# Modelling project failure and its mitigation in a time-stamped network of interrelated tasks

Christos Ellinas[1,*], Naoki Masuda[1,2,*]

[1] *Department of Engineering Mathematics, Merchant Ventures Building, University of Bristol, Woodland Road, Clifton, Bristol, BS8 1UB, United Kingdom*

[2] *Faculty of Management and Economics, Dalian University of Technology, No. 2 Linggong Road, Ganjingzi District, Dalian City, Liaoning Province, 116024, China*

[*] Corresponding authors

E-mail: ce12183@bristol.ac.uk, naoki.masuda@bristol.ac.uk

## Abstract

Resolving major societal challenges, such as stagnated economic growth or wasted resources, heavily relies on successful project delivery. However, projects are notoriously hard to deliver successfully, partly due to their interconnected nature which makes them prone to cascading failures. We deploy a model of cascading failure to temporal network data obtained from an engineering project, where tasks constituting the entire project and inter-dependencies between tasks correspond to time-stamped nodes and edges, respectively. We numerically evaluate the performance of six strategies to mitigate cascading failures. It is assumed that increased time between a pair of inter-connected tasks acts as a buffer, preventing a failure to propagate from one task to another. We show that, in a majority of cases that we explored, temporal properties of the activities (i.e., start and end date of each task in the project) are more relevant than their structural properties (i.e., out-degree and the size of the out-component of the task) to preventing large-scale cascading failures. Our results suggest potential importance of changing timings of tasks, apart from the static structure of the same network of tasks, for containing project failure.

## Introduction

Project-based processes are central in resolving major societal challenges[1], from accelerating economic growth (e.g., delivering infrastructure projects[2,3]) to fostering public resilience (e.g., mobilising resources in response to a natural hazard[4,5]). As an example, World Bank data from 2009 indicate that 22% of the world's gross domestic products, which is equivalent to approximately $48 trillion, relies almost entirely on project-based delivery mechanisms.[6] Successfully delivering projects is a non-trivial task, partly due to the interdependent nature of tasks composing a project.[7] A 2004 report by PricewaterhouseCoopers concluded that out of 10,640 projects reviewed in 30 countries and across a variety of industries, with a total value of approximately €5 billion, only 254 were successfully delivered.[8] A 2011 report concluded that out of 1,417 IT projects reviewed, 236 projects experienced cost overruns of at least 200% and the delivery of these projects was delayed by almost 70% in time.[9] Similar findings have also been reported elsewhere.[10,11] Because project scales are predicted to increase in the future (e.g., 1.5-2.5% annual growth in value over the past century[12]), the implications of project failure are expected to increase even further.

Research into understanding project failure can be broadly classified into two distinct, yet complementary, strands.[13] Several studies focus on mapping the sociological factors that contribute to project failure (e.g., importance of leadership[14,15] and corporate environment[16]). However, this strand of work is generally associated with a multitude of biases such as recollection bias (i.e., information bias in which recalled information is inaccurate) and self-report bias (i.e., behavioural bias in which participants over-report positive results), which challenges the integration of their findings to develop mitigation strategies against project failure.[13] A second approach relies on computational methods[17,18] that model the conditions of project failure, from lacking a 'healthy' organisational culture[19] to the propensity of wastefully repeating certain tasks.[20]

In computational approaches, a project is typically viewed as a directed acyclic graph and often called activity network[21], in which time-stamped nodes represent scheduled tasks (Figure 1). Directed edges between two nodes model functional dependencies between the two tasks. For example, a directed edge from node $i$ to node $j$ indicates that task $i$ must be completed before task $j$ begins. Because tasks are time-stamped, an activity network can be regarded as a temporal network.[22-24]

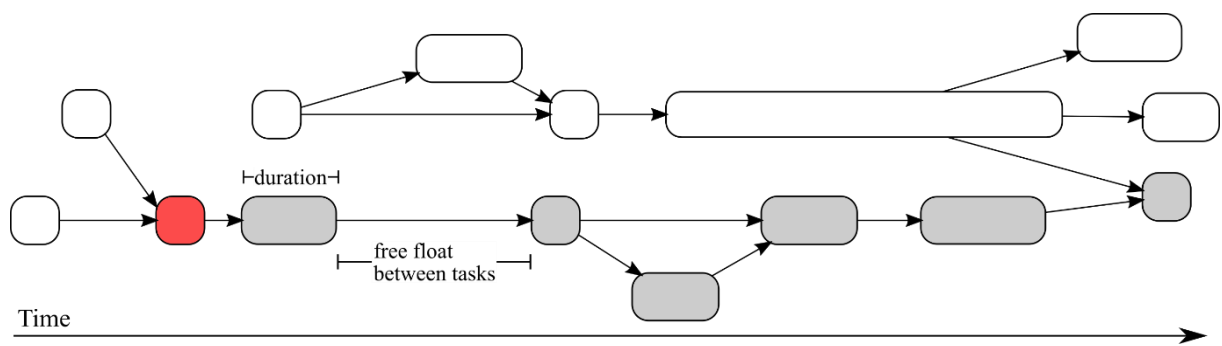

**Figure 1: Schematic of an activity network. A rounded rectangle represents a node (i.e., task). The grey rounded rectangles represent the tasks that may fail in response to a failure of the seed node shown in red.**

Early work on project failure focused on a particular failure scenario – delay propagation.[25,26] Under this scenario, a delay in the completion of a single task can propagate throughout the activity network, eventually delaying the entire project. A similar scenario has been examined in the context of delay propagation in airport networks.[27-29] Subsequent work on project failure has focussed on an alternative failure scenario in which changes in task specifications can trigger substantial rework in subsequent, downstream tasks and similarly affect the delivery of the overall project. In this case, a relatively minor change in the specifications of a single task can propagate across an entire project, severely affecting the overall project performance.[30-32]

These scenarios of project failure, in which a delay or a change in the task specification that occurs in a single task propagate across the overall project, seem to be an exemplary type of a cascading failure on activity networks. By cascading failure we refer to iterative processes in which a single failure leads to subsequent failures, which can potentially lead to a system-wide failure[33,34]. Past studies attributed a diverse set of system-wide failures to cascading failures, including financial systemic risk[35,36], the spread of misinformation[37,38], and power blackouts[39,40]. Along this line, our recent studies tackled long-lasting project-management challenges using network analysis (e.g., assessing potential of conflict between sub-contractors)[41] and associated certain project features with heightened vulnerability to cascading failures[42]. However, both studies did not aim to provide specific mitigation strategies with which to contain failure cascades, which is the main focus of the present study.

Robustness against cascading failures in networks can be engineered via structural or temporal mitigation schemes. Structural mitigation can be deployed when the structure of the network can be changed. For example, in power grids, one can modify the network structure to discourage the onset of large-scale cascades e.g., by introducing network modules or purposefully fragmenting the network before a cascade happens.[43,44] However, some network systems that are susceptible to cascading failures may not accommodate structural mitigation. In this situation, temporal mitigation, i.e., changing the timing of nodes or edges without changing the static network structure, may be deployed without compromising the function of the system. For example, in air traffic networks where nodes and edges are airports and flights respectively, delaying flights is probably much easier than changing the destination of the flights as a preventive measure against cascading failures. One can implement a temporal mitigation scheme if nodes or edges have timestamps that are relatively flexible. In air traffic networks, every edge has a timestamp that corresponds to the flight's scheduled departure time. An addition of a buffer between consecutive flights that depart from the same airport, which is a temporal mitigation strategy, may be able to suppress the likelihood of a small delay causing a system-wide delay.[27,29]

In project management contexts, deploying structural mitigation in activity networks is not often feasible because a directed edge from task $i$ (e.g., designing a structural column for a building) to task $j$ (e.g., manufacturing that column) indicates that task $i$'s output is necessary for starting task $j$, and therefore cannot be amended. As such, our focus here is on temporal as opposed to structural mitigation.

In the present study, we develop a simple cascading model which naturally lends itself to temporal active mitigation schemes. In other words, once a failure cascade is triggered, we deploy a mitigation scheme that modifies the start time of downstream tasks with the aim of suppressing cascading failures. To this end, we introduce a temporal element to the dynamics of these cascades by incorporating the impact of free float (i.e., the time between completion of task and the start of an immediately downstream task[45]) to a popular independent cascade model[46]. We then implement this model on an empirical activity network and numerically evaluate the performance of six mitigation schemes.

**Results**

The dataset is composed of a set of interconnected tasks that need to be completed for a commercial product in the area of defence to be delivered (see also Methods). The mean in- and out-degree of a task, which is regarded as a node of the directed temporal network, is equal to 1.69. The in-degree has standard deviation 4.45 and ranges from 0 to 90. A total of 111 nodes out of the 723 nodes has an in-degree of 0; these tasks are located in the most upstream position in the network, and starting any of these tasks does not need any other task to be completed beforehand. The out-degree has standard deviation 2.82 and ranges from 0 to 52. A total of 32 nodes has an out-degree of 0; these tasks are located in the most downstream position in the network, and failure of any of these tasks does not cause a cascading failure. The in- and out-degree obeys somewhat long-tailed distributions (Figure 2(a)), as is evidenced by their relatively large standard deviations as compared to the mean. The inter-event time has the mean equal to 141.4 days, standard deviation 169.5 days, and ranges from 0 to 670 days. The distribution of inter-event times is shown in Figure 2(b). The duration of task has the mean equal to 62.1 days, standard deviation 112.5 days, and ranges from 1 to 647 days. The distribution of the duration of tasks is shown in Figure 2(c). As time progresses, tasks are completed; the fraction of completed tasks by day is shown in Figure 2(d). The dataset of the temporal network of tasks, including the start time and end time of each task, is provided as Supplementary Information and available online (see Data Availability section).

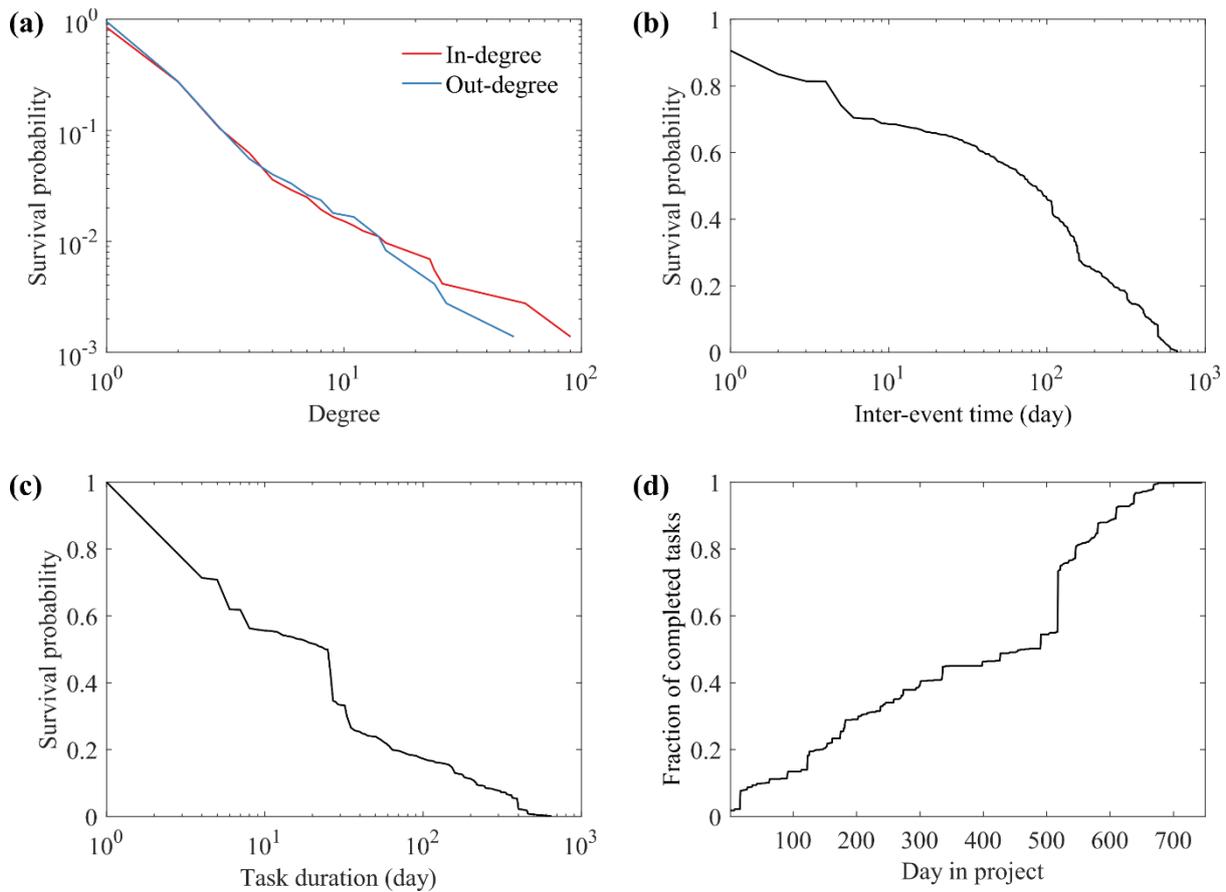

Figure 2: Distributions of basic properties of the temporal network of tasks. (a) Survival probability (i.e., probability that the degree is larger than or equal to a specified value) of the in- and out-degrees of the node. (b) Survival probability of the inter-event time. (c) Survival probability of the task duration. (d) Fraction of tasks that have been completed by day $t$, plotted against $t$.

To model cascading of failures, we introduce a variation of the independent cascade model[46]. In our model, the probability that a failure propagates from an affected node $i$ to a non-affected downstream neighbour node of node $i$, denoted by $j$, is a function of the probability that the failure of node $i$ causes the failure of node $j$ when there is no free float between the two nodes, which we denote by $q_0$, and the free float between the two nodes, which we denote by $\tau_{ij}$. We assume that the probability that the failure of node $i$ causes the failure of node $j$ decreases as $\tau_{ij}$ increases because a larger $\tau_{ij}$ indicates that more time is available for containing the effect of task $i$'s failure on its downstream neighbours. Another parameter $\tilde{\tau}$ controls the impact of the free float, $\tau_{ij}$, on the probability that the failure of node $i$ causes the failure of node $j$. By definition, a large $\tilde{\tau}$ value yields a small probability that the failure of node $i$ causes the failure of node $j$.

We evaluate the performance of the six mitigation schemes in terms of their ability of containing cascading failures. These mitigation schemes attempt to increase $\tau_{ij}$ for some $i$ and $j$ to reduce the probability that a failure cascades. Our focus is on the impact of the parameters that control the cascading dynamics ($q_0$ and $\tilde{\tau}$) and the fraction of the tasks to be postponed ($\gamma$).

With a mitigation scheme, we postpone some of the tasks located downstream to the seed node that has failed (Figure 3; also see the grey nodes in Figure 1). Doing so increases some of the inter-event times in the nodes belonging to the out-component of the seed node (i.e., the nodes downstream to the seed node). Therefore, a mitigation scheme is expected to reduce the probability that the failure propagates. Precisely, the fraction of the nodes in the out-component of the seed node for which we postpone the start time is denoted by $\gamma \in [0,1]$. A mitigation scheme is ranking of nodes in the out-component of the seed node. We sequentially postpone a fraction $\gamma$ of these nodes in descending order of the rank. When postponing each task $i$ sequentially, we postpone it as much as possible under the following two conditions. First, adjacent tasks must not overlap. In other words, the end date of task $i$ must not exceed the start date of any task $j$ that needs completion of task $i$. Second, the overall project duration must not be extended. In other words, the end date of task $i$ must not exceed the original delivery date of the project.

We test six mitigation schemes, in which nodes to be mitigated are ranked based on either the (i) out-degree, (ii) size of out-component (i.e., the number of nodes that are reachable from the node in question), (iii) duration of the task, (iv) start date of the task, (v) end date of the task or (vi) at random. For example, consider the network shown in the upper part of Figure 3(a) and assume that node $v_1$ fails. The subscript attached to the nodes in the figure represents the ranking in terms of the out-degree. The figure indicates that node $v_3$ is the first node to be mitigated (i.e., postponed). The amount of maximum postponement that can be applied to node $v_3$ is constrained by the start date of its immediate neighbour, node $v_4$. Therefore, we postpone node $v_3$ such that its new end date is equal to the start date of node $v_4$ (the network shown in the lower part of Figure 3(a)). Similarly, node $v_5$ is postponed such that its new end date is equal to the start date of node $v_6$. The same procedure is applied to node $v_2$ and then to node $v_6$. Note that postponing node $v_5$ makes the inter-event time between node $v_5$ and node $v_6$ equal to zero. However, postponing node $v_6$ subsequently increases the same inter-event time. We do not postpone the remaining two tasks with the lowest out-degrees, i.e., $v_4$ and $v_7$, because the fraction of the mitigated nodes, denoted by $\gamma$, is set to 0.67 for illustration purposes, such that only four out of the six nodes downstream to node $v_1$ can be mitigated. Implementation of three other mitigation schemes on the same network and the same $\gamma$ value is schematically shown in Figures 3(b)-3(d).

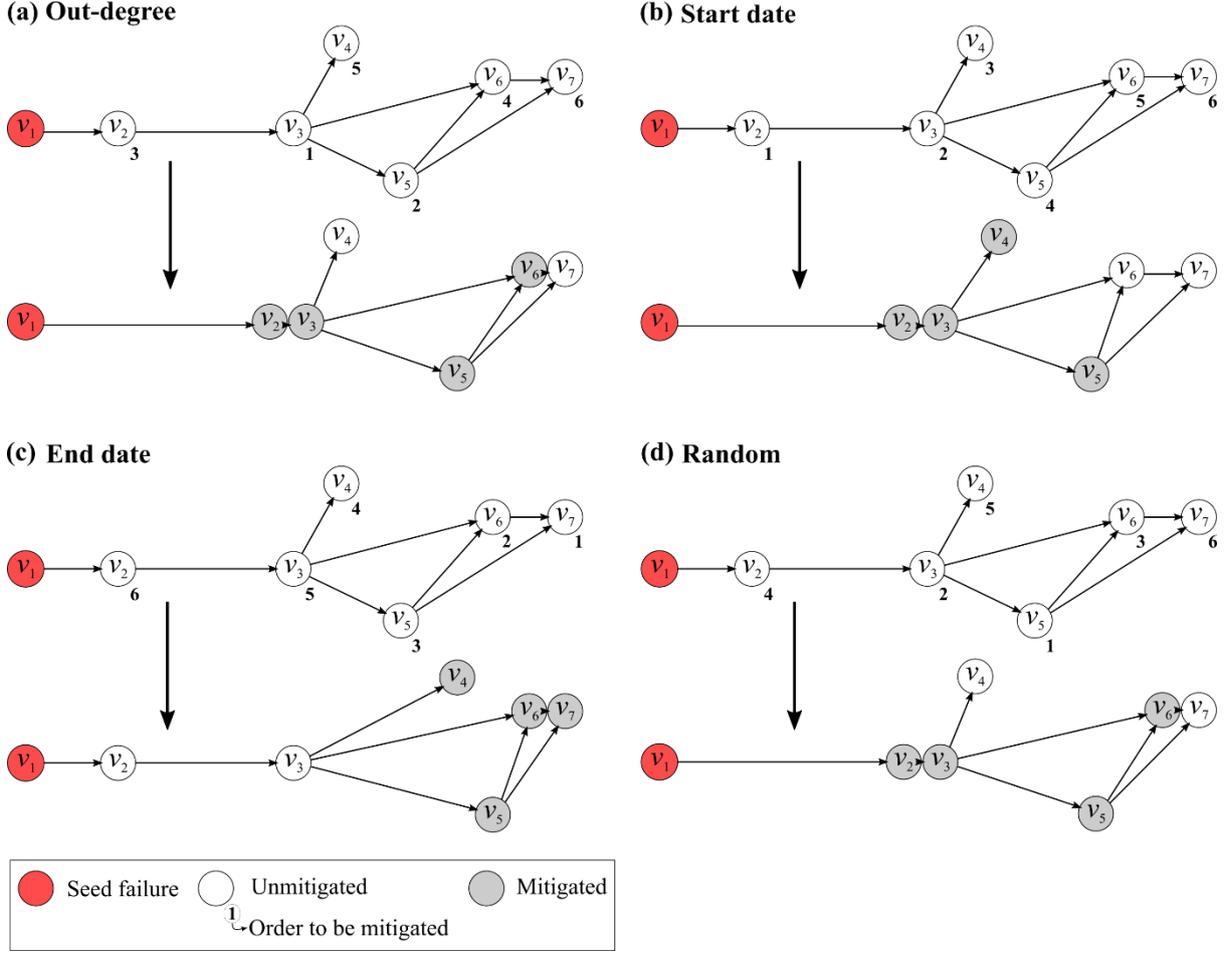

**Figure 3:** An example illustrating the four mitigation schemes: (a) out-degree, (b) start date, (c) end date and (d) random. For each mitigation scheme, the top and bottom panels correspond to before and after the mitigation, respectively. Every node is ranked (subscript index) and postponed in that order. The tie is broken uniformly randomly. In all examples, we set $\gamma = 0.67$ such that four out of the six tasks are mitigated.

When $\gamma = 1$, the mitigation scheme based on the end date of the task outperforms the other five mitigation schemes. This is the case in terms of both performance measures $R_1$ (Figure 4) and $R_2$ (Figure 5). Quantities $R_1$ and $R_2$ measure the relative and absolute reduction in the cascade size by a mitigation scheme (see Methods for the definitions). These figures also show that, apart from the mitigation scheme based on the end date of the task, the random mitigation scheme outperforms the other four mitigation schemes. The relative ranking of the six mitigation schemes is consistent in the whole range of $q_0$ and $\tilde{\tau} \in [1, 10, 10^2, 10^3]$, except for $\tilde{\tau} = 10^3$, where there are some rank changes presumably due to random fluctuations. Note that as $\tilde{\tau}$ tends large ($\tilde{\tau} \geq 10^3$), $p_{ij}$ is approximately equal to $q_0$ regardless of the size of $\tau_{ij}$ and regardless of the mitigation scheme. Therefore, $R_1$ and $R_2$ converge to 1 for any $q_0$ as $\tilde{\tau}$ increases (see Supplementary Figure 1 for numerical results with $\tilde{\tau} = 10^4$ and $\tilde{\tau} = 10^5$).

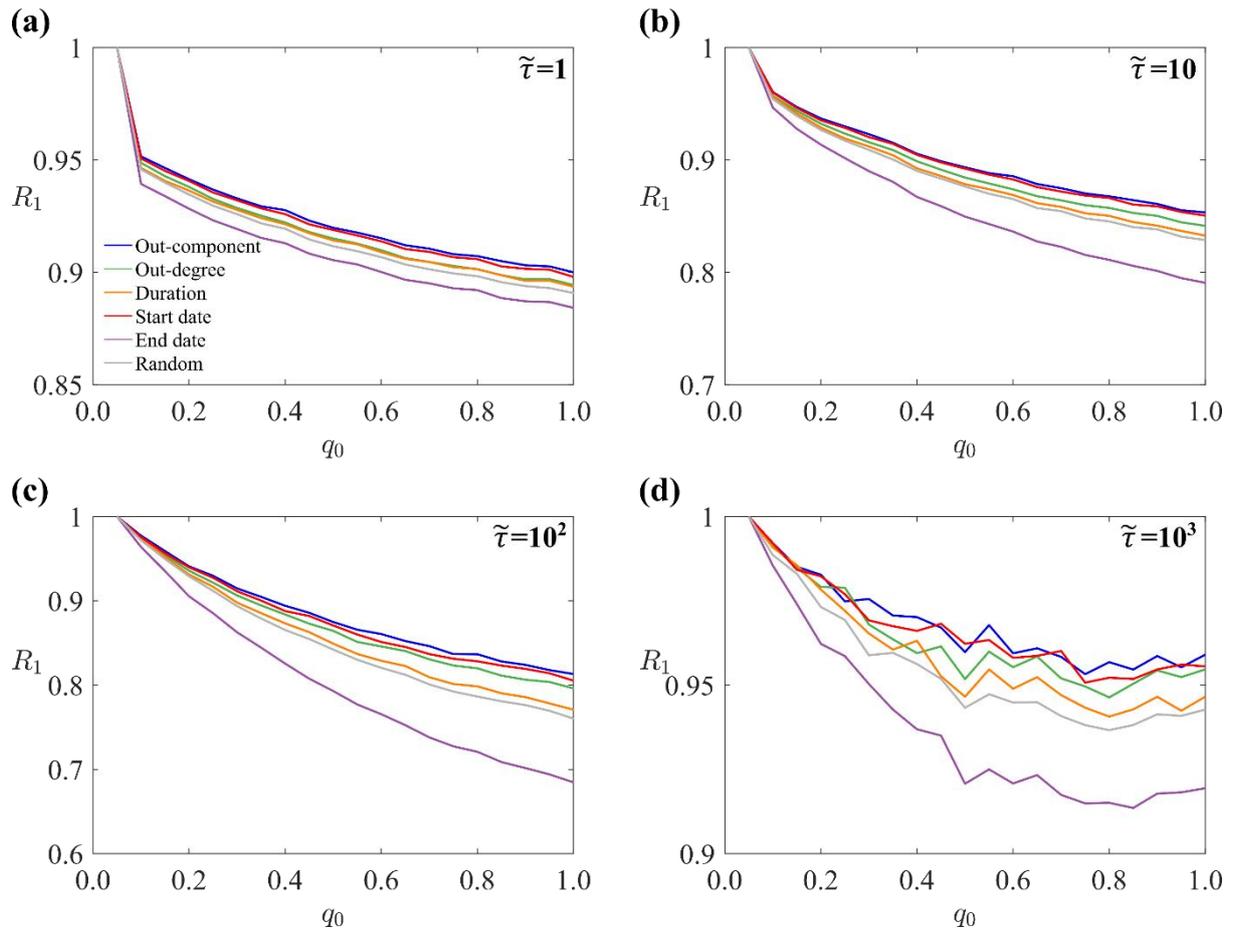

**Figure 4: Performance of the six mitigation schemes in terms of $R_1$, as a function of $q_0$.** (a) $\tilde{\tau} = 1$. (b) $\tilde{\tau} = 10$. (c) $\tilde{\tau} = 10^2$. (d) $\tilde{\tau} = 10^3$. We set $\gamma = 1$.

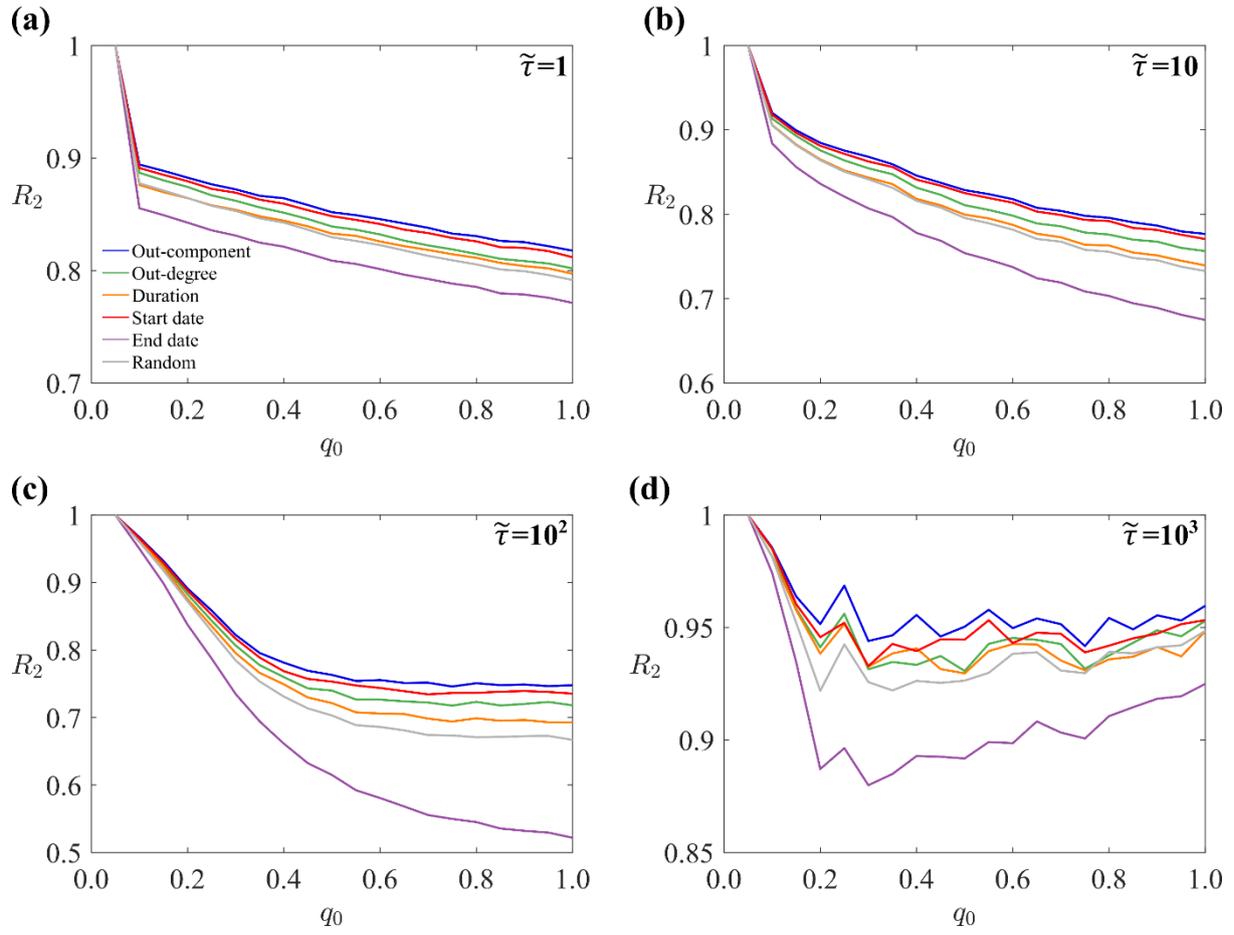

**Figure 5: Performance of the six mitigation schemes in terms of $R_2$, as a function of $q_0$.** (a) $\tilde{\tau} = 1$. (b) $\tilde{\tau} = 10$. (c) $\tilde{\tau} = 10^2$. (d) $\tilde{\tau} = 10^3$. We set $\gamma = 1$.

To investigate the entire parameter space where the fraction of mitigated nodes, $\gamma$, is also varied, we identified the mitigation scheme that was the most efficient, i.e., that yielding the smallest value of $R_1$ and $R_2$, when we varied $q_0$, $\tilde{\tau}$ and $\gamma$. The results in terms of $R_1$ are shown in Figure 6. When there is little variation between the best and worst performing schemes (less than 1%; arbitrarily chosen; white regions labelled 'Unspecified' in Figure 6), we argue that no best mitigation scheme exists. Figure 6 reveals two parameter regimes. First, when $\gamma \geq 0.8$, the mitigation scheme based on either the out-degree, duration, end date or at random performs the best, depending on the specific combination of $\gamma$ and $q_0$ values. As $\tilde{\tau}$ increases from 1 to $10^3$, the mitigation scheme based on the end date tends to be consistently the best in this parameter regime (Figure 6(d)). Second, when $\gamma < 0.8$, the mitigation scheme based on the start date tends to be the best performing mitigation scheme across the entire range of $q_0$ and $\tilde{\tau}$. The results in terms of $R_2$ (Supplementary Figure 2) are similar to those in terms of $R_1$ (Figure 6).

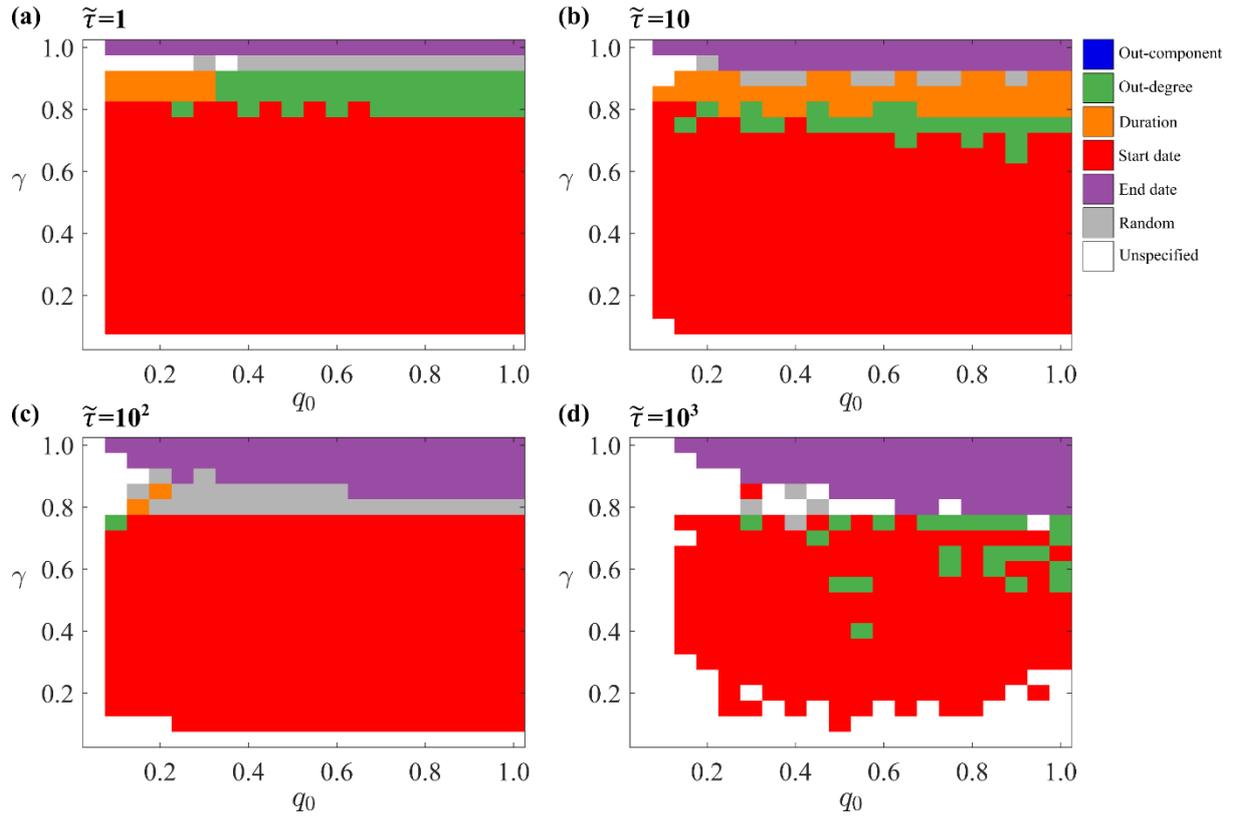

**Figure 6: Best performing mitigation scheme in terms of $R_1$ in the parameter space spanned by $q_0$ and $\gamma$.** (a) $\tilde{\tau} = 1$. (b) $\tilde{\tau} = 10$. (c) $\tilde{\tau} = 10^2$. (d) $\tilde{\tau} = 10^3$.

**Discussion**

We modelled project failures as cascading failures on networks composed of tasks constituting the project. The model incorporates both structural and temporal features of activity networks of projects. We implemented six mitigation schemes by postponing a fraction $\gamma$ of tasks downstream to the task that has failed. When one was allowed to postpone all the tasks downstream to the task that has failed, our numerical results indicated that it was more efficient to prioritise task mitigation according to the end date of each task than the other five mitigation schemes. When one was allowed to postpone a relatively small fraction of tasks, it was generally more efficient to postpone tasks based on their start date. Some additional cases existed where the mitigation scheme based on the out-degree or duration of the task was the most efficient. Specifically, when $\gamma$ is large, either the mitigation scheme based on the out-degree, that based on the duration, or that based on the end-date was the best. These numerical results suggest that, in a majority of the parameter region that we have explored, temporal features of the tasks, such as the duration, start date, and end date of the task, may be as important as structural features such as the out-degree of the task, for preventing large-scale cascading failures of projects.

The present modelling framework has limitations. First, our analysis has focused only on the benefits of deploying mitigation in the form of postponing the start date of tasks. However, postponing tasks may increase the number of active tasks on particular days, which is generally associated with poor project performance due to an increased cost or decreased quality.[20,47,48] Future work should consider this drawback in conjunction with the benefits potentially gained through the mitigation mechanisms proposed in the present study. Second, from a methodological standpoint, our approach is limited by the single pass in which mitigation is applied to tasks. Consider the example shown in Figure 3(a), in which node $v_5$ was postponed before node $v_6$ was. In this case, the amount of postponement is constrained by the start date of node $v_6$. Postponing node $v_6$ at a later stage opens up the opportunity for node $v_5$ to be further postponed, which is currently not exploited. One can exploit this opportunity to explore further improvements in mitigation efficiency. Third, a mitigation scheme can be classified into passive and active. In a passive mitigation scheme, one modifies the structure or time stamps of the activity network before a cascade is possibly seeded. In contrast, in an active mitigation scheme, one modifies the activity network while a cascade is progressing. In the present study, we focused on active mitigation schemes. Carefully planning the start time of each task, given, for example, the network structure and the possibility of different tasks to fail with different probabilities, may consist in a plausible passive mitigation scheme on the activity network. This topic also warrants future work.

Despite these and other possible limitations, we believe that the present modelling framework serves as a stepping stone for future work for exploring whether causal relationships exist between structural and/or temporal features of temporal networks of tasks and mitigation efficiency.

**Methods**

*Data*

The data set corresponds to a list of $N = 723$ planned activities, which we refer to as tasks, that need to be completed for a commercial product in the area of defence to be delivered. The overall duration of the project is 745 days. Each task has a scheduled start and end date (and hence the duration). The resolution of the time is a day. The dependency between a pair of tasks is represented by a directed edge. There are 1,220 directed edges in total. The directed edge from task $i$ to $j$, denoted by $e_{ij} \in E$, indicates that the output of task $i$, such as information or a physical artefact (i.e., product), is an input to task $j$. A directed edge form task $i$ to task $j$ implies that task $i$ must be completed before task $j$ starts. Therefore, task $j$ can start only after all tasks that send a directed edge to task $j$ have been completed. The 723 activities as nodes and the 1,220 edges define an activity network, which is a time-stamped directed acyclic graph. The number of predecessors and successors of each task is equal to the task's in-degree and out-degree, respectively.

The free float between task $i$ and $j$ is defined as the difference, in days, between the completion of task $i$ and the start of task $j$ [45]. The free float is equivalent to a widely used term, inter-event time.[22-24] We denote the free float between $i$ and $j$ as $\tau_{ij}$.

*Cascading failure model*

We use a discrete-time cascading failure model with binary states of the node, which is analogous to the independent cascade model[46] and other cascade-failure models[33]. The final state of node $j$ ($1 \leq j \leq N$) is denoted by $s_j \in [0,1]$, where '0' and '1' correspond to the non-affected and affected state, respectively. We start the cascade dynamics from an initial condition, where one seed node (which can be any node) is in state 1 and all the other $N - 1$ nodes are in state 0. During the cascade dynamics, node $j$ may irreversibly switch from state 0 to state 1 if node $j$ has at least one upstream neighbour that is in state 1.

We determine the final state of each node (and hence the final cascade size) by marking the nodes one by one as follows. Note that the results do not depend on the order of marking the nodes. Initially, the seed node is the only marked node (i.e., finalised to state 1) in the network. During the course of the following procedure, all nodes that are yet to be marked have state 0. Marked nodes have state either 0 or 1. In each round, we pick an unmarked node $j$ whose all upstream neighbours have been marked. The first node to be marked after the seed node is a node that does not have any upstream neighbour (i.e., in-degree equal to 0) or a node that has the seed node as the only upstream neighbour. In the former case, the node selected for marking is typically a task that starts the earliest in time. To determine the final state of node $j$ (i.e., to mark node $j$), we assume that the failure of each upstream neighbour of node $j$, referred to as node $i$, independently causes node $j$ to fail with probability $p_{ij}$. Then, we set the final state of node $j$ to 1 with probability

$$P_j = 1 - \prod_{i;\ e_{ij} \in E} [(1 - s_i) + s_i(1 - p_{ij})], \tag{1}$$

where $E$ is the set of edges. Otherwise, we set the final state of node $j$ to 0. In eq. (1), the product term is the probability that node $j$ does not fail, and each factor in the product is the probability that node $i$ does not cause the failure of node $j$. If $s_i = 0$, this probability is equal to 1. If $s_i = 1$, this probability is equal to $1 - p_{ij}$. Once the state of node $j$ is determined in this manner, we mark node $j$ and select a next unmarked node such that all its upstream neighbours have been marked.

To set the value of $p_{ij}$, we consider the impact of time between the completion of task $i$ and start of task $j$ (i.e., inter-event time, denoted by $\tau_{ij}$). We reason that a large inter-event time reduces the probability that task $i$'s failure propagates to task $j$, because an inter-event time gives spare time to dispense appropriate resources to mitigate the effect of task $i$'s failure.[49,50] Reducing inter-event times has been suggested to increase the risk of failure propagation as well.[51] Therefore, we assume that

$$p_{ij} = q_0 \exp\left(-\frac{\tau_{ij}}{\tilde{\tau}}\right), \tag{2}$$

where $q_0 \in [0,1]$ and $\tilde{\tau}(> 0)$ are parameters. Parameter $q_0$ is the probability that task $j$ fails if task $i$ does and there is no spare time between the two tasks, i.e., $\tau_{ij} = 0$. Equation (2) indicates that if the two tasks are far apart in time, it is not likely that failure of one task triggers failure of a successor task. Parameter $\tilde{\tau}$ controls the strength of this time dependence.

*Mitigation schemes*

By the construction of our model, increasing an inter-event time reduces the probability that failure propagates from a task to another. Therefore, postponing the start of a downstream task is expected to reduce the probability of its failure. We use this mechanism as a mitigation scheme. A mitigation scheme has to respect the end date of the entire project; no task can be postponed beyond the delivery date of the entire project. Furthermore, any downstream neighbour of task $j$ is only allowed to start after task $j$ has been completed. Therefore, the extent of postponing task $j$ is further constrained by the start date of its downstream neighbours. Note that we allow the end date of task $j$ to coincide with the start date of its downstream neighbour, in which case the inter-event time is equal to zero.

The mitigation scheme is implemented as follows. Once seed node $i$ fails, all nodes reachable from node $i$ along a directed path (i.e., nodes belonging to the out-component of node $i$), which can fail, are rank ordered based on the node's score. The score of these nodes is equal to one of the following six quantities: out-degree, size of the out-component (i.e., the number of nodes that are reachable from the node to be scored), duration of the task, start date of the task, end date of the task, or an entirely randomly drawn value. When multiple nodes have identical scores, we break the tie by ranking the nodes having the same score in a uniformly random order.

We denote by $\tilde{V}$ the rank-ordered set of the nodes downstream to node $i$. In the example shown in Figure 3a, in which the rank is determined according to the out-degree of the task, we obtain $\tilde{V} = \{v_3, v_5, v_2, v_6, v_4, v_7\}$. Parameter $\gamma \in [0,1]$ specifies the fraction of nodes in $\tilde{V}$ that are to be mitigated. In Figure 3(a), we set $\gamma = 0.67$. Therefore, the four highest-ranked nodes out of the six nodes, i.e., $v_3, v_5, v_2$ and $v_6$, are mitigated. Node $v_3$ is first postponed until its end date coincides with the start date of its downstream neighbour $v_4$. Next, the same postponement process is applied to node $v_5$, node $v_2$ and then node $v_6$. The temporal network after the mitigation is shown in the lower part of Figure 3(a).

In the example shown in Figure 3(a), the probability that the failure of node $v_1$ propagates to node $v_2$ has been reduced because $\tau_{v_1 v_2}$ has been increased. This is a positive effect of mitigation that we have intended. However, in the same example, the probability that the failure of node $v_2$ propagates to node $v_3$ has been increased compared to the case of the unmitigated activity network, because the mitigation has decreased $\tau_{v_2 v_3}$. This is a negative effect of mitigation that we have not intended.

*Performance measures for mitigation schemes, $R_1$ and $R_2$*

We measure the performance of each mitigation scheme in terms of two quantities. The first quantity, denoted by $R_1$, is defined as the cascade size that stems from a seed node when the mitigation scheme is implemented, divided by the cascade size when there is no mitigation, averaged over all seed nodes. Quantity $R_1$ captures the relative impact of mitigation in the sense that the contribution of mitigating a large cascade is equivalent to that of mitigating a small cascade. The second quantity, denoted by $R_2$, is defined as the cascade size averaged over all seed nodes when the mitigation is applied, which is then divided by the cascade size averaged over all seed nodes when no mitigation is applied. Quantity $R_2$ captures the absolute impact of mitigation in the sense that mitigating a large cascade is considered to be more valuable than mitigating a small cascade. A small $R_1$ or $R_2$ value indicates that the mitigation scheme is efficient.

For the given values of $q_0$, $\tilde{\tau}$, $\gamma$, and the given seed node, we ran the cascading dynamics 100 times (except for Supplementary Figure 1, for which we ran the simulation 300 times). In the figures we show the average values of the observables over all runs.

**Data Availability**

The datasets generated during and/or analysed during the current study are available in the GitHub repository, https://github.com/naokimas/project_network.

**References**

1   Jensen, A., Thuesen, C. & Geraldi, J. The projectification of everything: Projects as a human condition. *Project Management Journal* **47**, 21-34; 10.1177/875697281604700303 (2016).

2   Grimsey, D. & Lewis, M. K. Evaluating the risks of public private partnerships for infrastructure projects. *International Journal of Project Management* **20**, 107-118; 10.1016/S0263-7863(00)00040-5 (2002).

3   Love, P. E., Davis, P. R., Chevis, R. & Edwards, D. J. Risk/reward compensation model for civil engineering infrastructure alliance projects. *Journal of Construction Engineering and Management* **137**, 127-136; 10.1061/(ASCE)CO.1943-7862.0000263 (2010).

4   Lin Moe, T. & Pathranarakul, P. An integrated approach to natural disaster management: Public project management and its critical success factors. *Disaster Prevention and Management* **15**, 396-413; 10.1108/09653560610669882 (2006).

5   Kusumasari, B., Alam, Q. & Siddiqui, K. Resource capability for local government in managing disaster. *Disaster Prevention and Management* **19**, 438-451; 10.1108/09653561011070367 (2010).

6   Scranton, P. Projects as a focus for historical analysis: surveying the landscape. *History and Technology* **30**, 354-373; 10.1080/07341512.2014.1003164 (2014).

7   Baccarini, D. The concept of project complexity—a review. *International Journal of Project Management* **14**, 201-204; 10.1016/0263-7863(95)00093-3 (1996).

8	PricewaterhouseCoopers. Boosting Business Performance through Programme and Project Management. Available at: https://www.mosaicprojects.com.au/PDF/PwC_PM_Survey_210604.pdf (2004) [Accessed on 15/10/2018].

9	Flyvbjerg, B. & Budzier, A. Why your IT project may be riskier than you think. *Harvard Business Review* **89**, 601-603 (2011).

10	Bar-Yam, Y. When systems engineering fails - toward complex systems engineering, *2003 IEEE International Conference on Systems, Man and Cybernetics. Conference Theme - System Security and Assurance, Washington DC, USA* **2**, 2021-2028; 10.1109/ICSMC.2003.1244709 (2003).

11	Flyvbjerg, B., Skamris holm, M. K. & Buhl, S. L. How common and how large are cost overruns in transport infrastructure projects? *Transport Reviews* **23**, 71-88; 10.1080/01441640309904 (2003).

12	Flyvbjerg, B. What you should know about megaprojects and why: An overview. *Project Management Journal* **45**, 6-19; 10.1002/pmj.21409 (2014).

13	Parvan, K., Rahmandad, H. & Haghani, A. Inter-phase feedbacks in construction projects. *Journal of Operations Management* **39**, 48-62; 10.1016/j.jom.2015.07.005 (2015).

14	Turner, J. R. & Müller, R. The project manager's leadership style as a success factor on projects: A literature review. *Project Management Journal* **36**, 49-61; 10.1177/875697280503600206 (2005).

15	Snowden, D. J. & Boone, M. E. A leader's framework for decision making. *Harvard Business Review* **85**, 68-76 (2007).

16	Schmidt, R., Lyytinen, K., Keil, M. & Cule, P. Identifying software project risks: An international Delphi study. *Journal of Management Information Systems* **17**, 5-36; 10.1080/07421222.2001.11045662 (2001).

17	Ashworth, M. J. & Carley, K. M. Can tools help unify organization theory? Perspectives on the state of computational modeling. *Computational and Mathematical Organization Theory* **13**, 89-111; 10.1007/s10588-006-9000-9; (2007).

18	Holme, P. & Liljeros, F. Mechanistic models in computational social science. *Frontiers in Physics* **3**, 78; 10.3389/fphy.2015.00078 (2015).

19	Ellinas, C., Allan, N. & Johansson, A. Dynamics of organizational culture: Individual beliefs vs. social conformity. *PLoS ONE* **12**, e0180193; 10.1371/journal.pone.0180193 (2017).

20	Smith, R. P. & Eppinger, S. D. Identifying controlling features of engineering design iteration. *Management Science* **43**, 276-293; 10.1287/mnsc.43.3.276 (1997).

21	Browning, T. R. & Ramasesh, R. V. A survey of activity network-based process models for managing product development projects. *Production and Operations Management* **16**, 217-240; 10.1111/j.1937-5956.2007.tb00177.x (2007).

22	Holme, P. & Saramäki, J. Temporal networks. *Physics Reports* **519**, 97-125; 10.1016/j.physrep.2012.03.001 (2012).

23	Holme, P. Modern temporal network theory: A colloquium. *European Physical Journal B* **88**, 234; 10.1140/epjb/e2015-60657-4 (2015).

24  Masuda, N. & Lambiotte, R. *A Guide to Temporal Networks*. Singapore: World Scientific (2016).

25  Kelley Jr, J. E. & Walker, M. R. Critical-path planning and scheduling, *IRE-AIEE-ACM Computer Conference*, 160-173 (1959).

26  Malcolm, D. G., Roseboom, J. H., Clark, C. E. & Fazar, W. Application of a technique for research and development program evaluation. *Operations Research* **7**, 646-669; 10.1287/opre.7.5.646 (1959).

27  AhmadBeygi, S., Cohn, A., Guan, Y. & Belobaba, P. Analysis of the potential for delay propagation in passenger airline networks. *Journal of Air Transport Management* **14**, 221-236; 10.1016/j.jairtraman.2008.04.010 (2008).

28  Fleurquin, P., Ramasco, J. J. & Eguiluz, V. M. Systemic delay propagation in the US airport network. *Scientific Reports* **3**, 1159; 10.1038/srep01159 (2013).

29  Ivanov, N., Netjasov, F., Jovanović, R., Starita, S. & Strauss, A. Air traffic flow management slot allocation to minimize propagated delay and improve airport slot adherence. *Transportation Research Part A* **95**, 183-197; 10.1016/j.tra.2016.11.010 (2017).

30  Terwiesch, C. & Loch, C. H. Managing the process of engineering change orders: The case of the climate control system in automobile development. *Journal of Product Innovation Management* **16**, 160-172; 10.1016/S0737-6782(98)00041-1 (1999).

31  Mihm, J., Loch, C. & Huchzermeier, A. Problem-solving oscillations in complex engineering projects. *Management Science* **49**, 733-750; 10.1287/mnsc.49.6.733.16021 (2003).

32  Sosa, M. E. Realizing the need for rework: From task interdependence to social networks. *Production and Operations Management* **23**, 1312-1331; 10.1111/poms.12005 (2014).

33  Lorenz, J., Battiston, S. & Schweitzer, F. Systemic risk in a unifying framework for cascading processes on networks. *European Physical Journal B* **71**, 441-460; 10.1140/epjb/e2009-00347-4 (2009).

34  Motter, A. E. & Yang, Y. The unfolding and control of network cascades. *Physics Today* **70**, 32-39; 10.1063/PT.3.3426 (2017).

35  Gai, P. & Kapadia, S. Contagion in financial networks. *Proceedings of the Royal Society A* **466**, 2401–2423; 10.1098/rspa.2009.0410 (2010).

36  Acemoglu, D., Ozdaglar, A. & Tahbaz-Salehi, A. Systemic risk and stability in financial networks. *American Economic Review* **105**, 564-608; 10.3386/w18727 (2015).

37  Vosoughi, S., Roy, D. & Aral, S. The spread of true and false news online. *Science* **359**, 1146-1151; 10.1126/science.aap9559 (2018).

38  Zhao, Z. et al. Fake news propagate differently from real news even at early stages of spreading. Preprint at https://arxiv.org/abs/1803.03443 (2018).

39  Dobson, I., Carreras, B. A., Lynch, V. E. & Newman, D. E. Complex systems analysis of series of blackouts: Cascading failure, critical points, and self-organization. *Chaos* **17**, 026103; 10.1063/1.2737822 (2007).

40  Yang, Y., Nishikawa, T. & Motter, A. E. Small vulnerable sets determine large network cascades in power grids. *Science* **358**, eaan3184; 10.1126/science.aan3184 (2017).

41    Ellinas, C., Allan, N. & Johansson, A. Project systemic risk: Application examples of a network model. *International Journal of Production Economics* **182**, 50-62; 10.1016/j.ijpe.2016.08.011 (2016).

42    Ellinas, C. The domino effect: An empirical exposition of systemic risk across project networks. *Production and Operations Management*; 10.1111/poms.12890 (2018).

43    Pagani, G. A. & Aiello, M. The power grid as a complex network: A survey. *Physica A* **392**, 2688-2700; 10.1016/j.physa.2013.01.023 (2013).

44    Guo, H., Zheng, C., Iu, H. H.-C. & Fernando, T. A critical review of cascading failure analysis and modeling of power system. *Renewable and Sustainable Energy Reviews* **80**, 9-22; 10.1016/j.rser.2017.05.206 (2017).

45    Elmaghraby, S. E. Activity nets: A guided tour through some recent developments. *European Journal of Operational Research* **82**, 383-408 (1995).

46    Goldenberg, J., Libai, B. & Muller, E. Talk of the network: A complex systems look at the underlying process of word-of-mouth. *Marketing Letters* **12**, 211-223; 10.1023/A:1011122126881 (2001).

47    Krishnan, V., Eppinger, S. D. & Whitney, D. E. A model-based framework to overlap product development activities. *Management Science* **43**, 437-451; 10.2307/2634558 (1997).

48    Cho, S.-H. & Eppinger, S. D. A simulation-based process model for managing complex design projects. *IEEE Transactions on Engineering Management* **52**, 316-328; 10.1109/TEM.2005.850722 (2005).

49    Buzna, L., Peters, K., Ammoser, H., Kühnert, C. & Helbing, D. Efficient response to cascading disaster spreading. *Physical Review E* **75**, 056107; 10.1103/PhysRevE.75.056107 (2007).

50    Ellinas, C., Allan, N., Durugbo, C. & Johansson, A. How robust is your project? From local failures to global catastrophes. *PLoS ONE* **10**, e0142469; 10.1371/journal.pone.0142469 (2015).

51    Project Management Institute, *A Guide to the Project Management Body of Knowledge*, Newton Square, Pennsylvania: Project Management Institute (2016).

**Acknowledgements**

CE acknowledges the support provided by ENCORE+ and EPSRC Doctoral Prize fellowship. NM acknowledges the support provided through JST, CREST, Grant Number JPMJCR1304. CE is particularly grateful to Thales UK for providing the data.

**Additional Information**

*Author contributions*

CE and NM designed the research. CE obtained and pre-processed the data, performed numerical simulations and analysed the results. CE and NM wrote and reviewed the manuscript.

*Competing financial interests*

The authors declare no competing financial interests.

# Supplementary Information for "Modelling project failure and its mitigation in a time-stamped network of interrelated tasks"

Christos Ellinas[1,*], Naoki Masuda[1,2,*]

[1] *Department of Engineering Mathematics, Merchant Ventures Building, University of Bristol, Woodland Road, Clifton, Bristol, BS8 1UB, United Kingdom*

[2]*Faculty of Management and Economics, Dalian University of Technology, No. 2 Linggong Road, Ganjingzi District, Dalian City, Liaoning Province, 116024, China*

*Corresponding authors

E-mail: ce12183@bristol.ac.uk, naoki.masuda@bristol.ac.uk

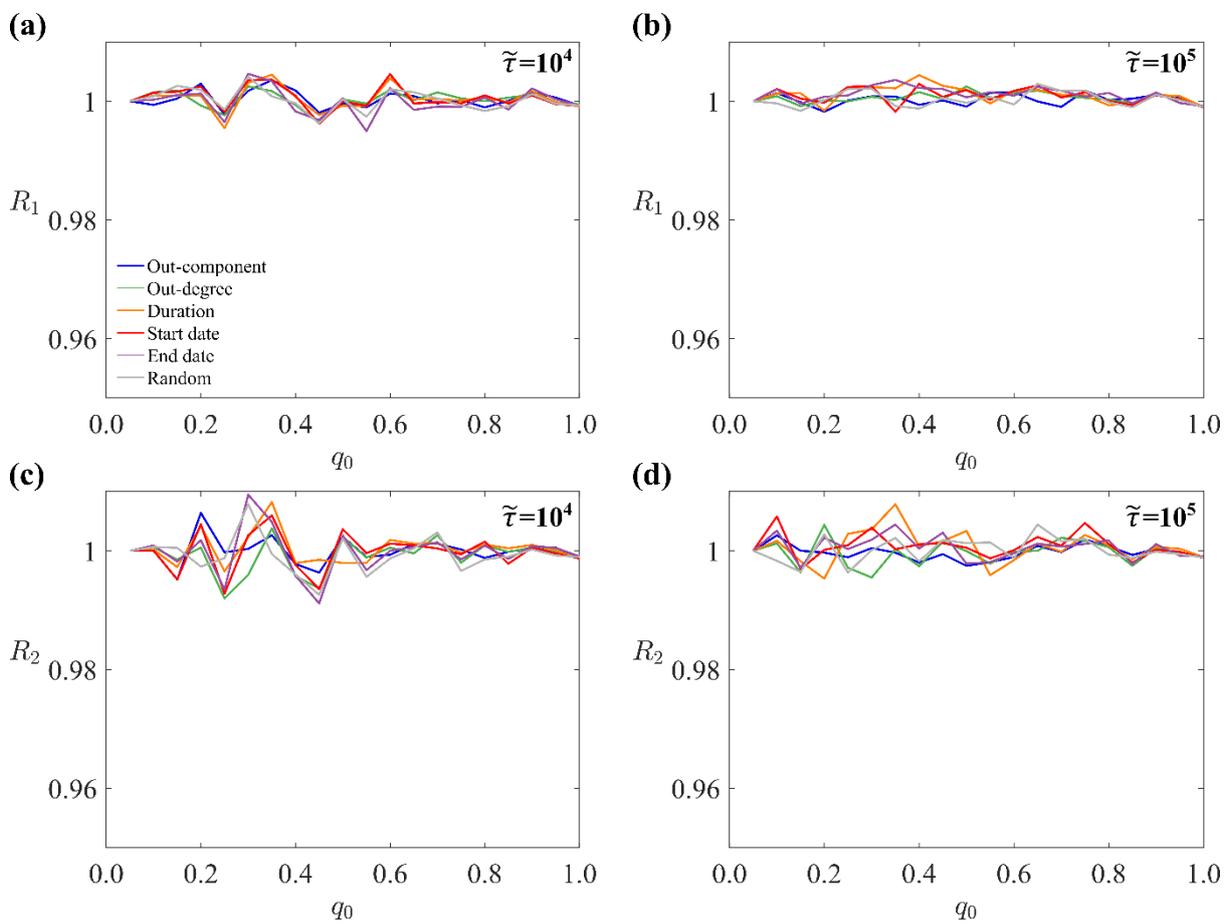

**Supplementary Figure 1: Performance of the six mitigation schemes for larger values of $\tilde{\tau}$.** (a) $R_1$ when $\tilde{\tau} = 10^4$. (b) $R_1$ when $\tilde{\tau} = 10^5$. (c) $R_2$ when $\tilde{\tau} = 10^4$. (d) $R_2$ when $\tilde{\tau} = 10^5$.

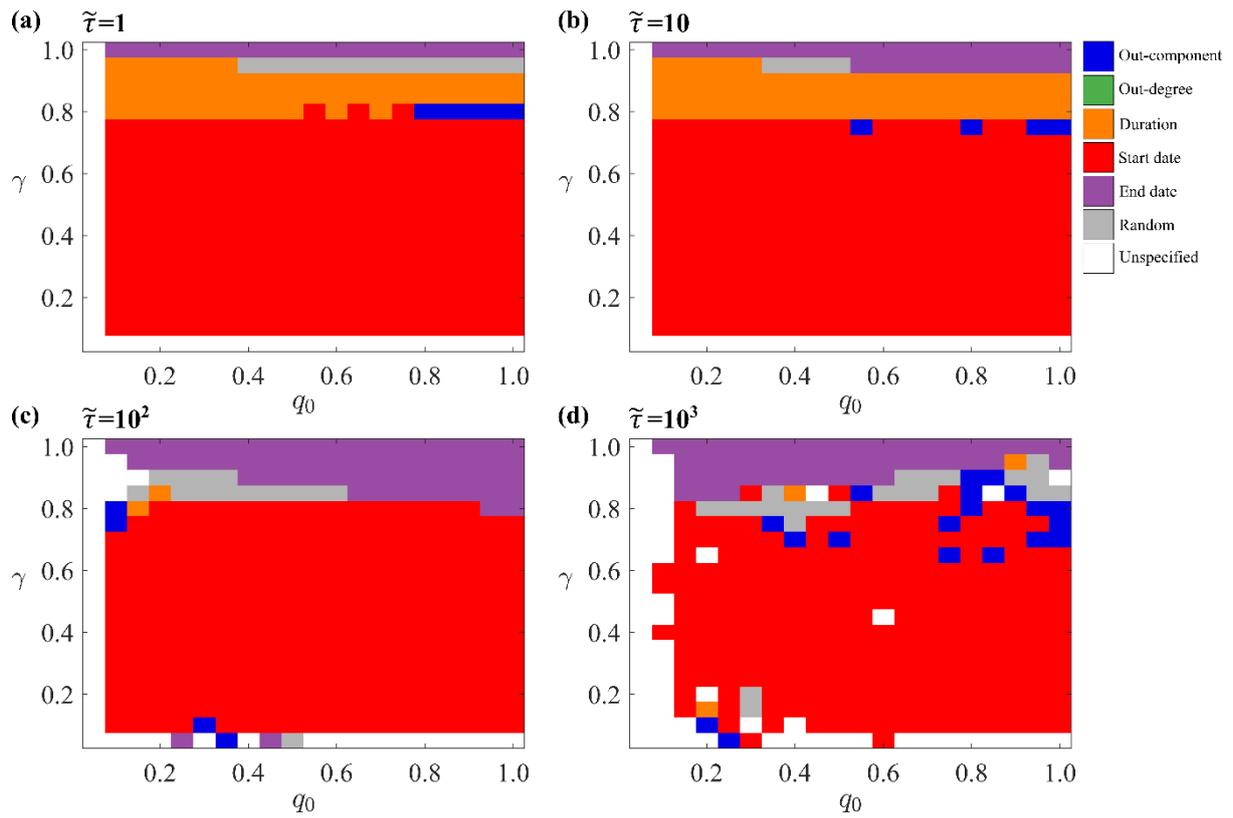

**Supplementary Figure 2: Best performing mitigation scheme in terms of $R_2$ in the parameter space spanned by $q_0$ and $\gamma$.** (a) $\tilde{\tau} = 1$. (b) $\tilde{\tau} =$. (c) $\tilde{\tau} = 10^2$. (d) $\tilde{\tau} = 10^3$.